\documentclass{article}		
\usepackage{latexsym}		

\begin{document}
\def\H2{$(---+)$}
\def\R4{$(+++-)$}
\def\Ed#1{{\bf E}_{(#1)}}
\def\bas#1{\gamma_{#1}}
\def\basis#1{{\gamma}^{#1}}
\def\bivtr#1#2{\gamma^{#1}\gamma^{#2}}
\def\trivtr#1#2#3{\gamma^{#1}\gamma^{#2}\gamma^{#3}}
\def\quadvtr{\trivtr{1}{2}{3}\gamma^4}
\def\g5{\epsilon}
\def\basise#1{{\bf e}_{#1}}
\def\basisu#1{{\bf e}^{#1}}
\def\vector#1{\vec{\bf {#1}}}        
\def\wmp#1{{\tt #1}}
\def\voros#1{{\sf #1}}
\def\jv#1{{\sf #1}}
\def\U#1{\underline{#1}}
\def\Tau{\tau}
\def\til#1{{#1} \rightarrow -{#1}}
\def\tilt#1{$\til{#1}$}
\def\tilti#1{${#1} \rightarrow i\ {#1}$}
\def\norm#1{\parallel{#1}\parallel}
\def\BOXY{\Box}
\def\BAR#1{\overline{#1}}
\def\EQN#1{eq.\ (#1)}
\def\hbar{{\mathchar'26\mkern-9muh}}
\begin{titlepage}

\title{\Large\bf Should Metric Signature Matter in 
Clifford Algebra Formulations of Physical Theories?
\thanks{anonymous ftp://www.clifford.org/clf-alg/preprints/1995/pezz9502.latex}}

\author{{\Large\bf William M. Pezzaglia Jr.}
\thanks{Email: wpezzaglia@scuacc.scu.edu or billium@well.com}
\\ Department of Physics \\ Santa Clara University
\\ Santa Clara, CA 95053 \\
 \\ \and
{\Large\bf John J. Adams}
\thanks{Email: jadams@llnl.gov}
\\ P.O. Box 808, L-399 \\
Lawrence Livermore National Laboratory \\ Livermore, CA 94550 }

\maketitle
\thispagestyle{empty}

\begin{abstract}
Standard formulation is unable to distinguish between the \R4 and \H2
spacetime metric signatures.  However, the Clifford algebras associated
with each are inequivalent, {\bf R}(4) in the first case (real 4 by 4
matrices), {\bf H}(2) in the latter (quaternionic 2 by 2).  Multivector
reformulations of Dirac theory by various authors look quite inequivalent
pending the algebra assumed.  It is not clear if this is mere artifact,
or if there is a right/wrong choice as to which one describes reality.
However, recently it has been shown that one can map from one signature
to the other using a {\it tilt transformation}\cite{Lounesto93}.
The broader question is that if the universe is signature blind, then
perhaps a complete theory should be manifestly tilt covariant.
A generalized multivector wave equation
is proposed which is fully signature invariant in form, because it 
includes all the components of the algebra in the wavefunction
(instead of restricting it 
to half) as well as all the possibilities for interaction terms.

Summary of talk at the Special Session on {\it Octonions and Clifford
Algebras}, at the 1997 Spring Western Sectional Meeting of the American
Mathematical Society, Oregon State University, Corvallis, OR, 19-20
April 1997.
\end{abstract}

\end{titlepage}
\pagestyle{myheadings}

\section*{I. Introduction}
It is a well-known fact that physical predictions 
of relativity are independent of the absolute choice of
\R4 vs. \H2 metric signatures.  If this is generally true for 
\underline{all} physical laws, it would conversely follow that no
physical measurement can determine the absolute metric signature.
This is akin to the Einstein's cosmological metaprinciple: that one cannot
determine an \underline{absolute} inertial reference frame of the universe.
The truth of such broad statements can not be deduced or derived
from the accepted principles of physics, anymore than one can ``derive''
Einstein's weak equivalence principle (the requirement that
theories be locally Lorentz covariant).  The laws which \underline{all}
the laws of physics must obey, called {\it metalaws},
must be imposed at the onset.

Many of the current treatments have been concerned with the possibility
of the sign of the signature changing from one place to
another\cite{Embacher} or even changing from Lorentzian to 
Euclidean\cite{Carlini}.
At minimum this raises various
topological issues at the boundries between the regions, and perhaps
some exotic physics such as pair production\cite{Dray93}.
However, in our opinion it is premature to argue for physical
interpretation of phenomena at the local discontinuities of metric
change if the broader issues have not first been addressed.
Specifically we should be asking if a new metalaw should be stated:
{\it Physical laws must be globally invariant under the point transformation
\tilt{g_{\mu\nu}}}.  For example, in section II we consider the
behavior of non-relativistic classical electrodynamics under the change
of {\tt 3D} metric signature: $(+++)\rightarrow (---)$.
The observable physics appears to be completely signature blind,
however the  Gibbs vector algebra (n.b. the curl in Maxwell's 
equations) is not well suited to this type of abuse.
The alternate mathematical language of {\it
Clifford Geometric Algebra}\cite{Hest66} is better behaved, allowing for a
formulation of Maxwell's equations that is
completely {\it signature invariant in form}.  In section III however
we find that Quantum Mechanics \underline{cannot} be put into
signature invariant form.  Specifically the expectation values of
the physical observables cannot be made invariant unless the
wavefunction is allowed to change in form, as well as the wave
equation and Dirac spin algebra.

In contrast to standard formulation, there has been a plethora of claims
and results attributed to the \underline{absolute} choice
of metric signature in Clifford algebra ``Multivector'' formulations of physical
theories.  Hestenes\cite{Hestenes} in particular champions the \H2
choice while others[6-7] use the complement \R4.  It is a fact
that the algebraic structure of the Clifford algebra
for the two different signatures is completely inequivalent.
It follows that physical theories formulated with Clifford algebra are
therefore potentially inequivalent pending the underlying choice of signature.
In section IV we consider the dilemna:  Are the multivectorial formulations
of Dirac theory based upon the different metric signatures inequivalent
in their physical description {\it hence there is a testable
absolute right or wrong
one to match to reality}?  Or is there a metalaw
of {\it signature invariance}, which restricts the form of both such that the
differences are mere artifacts with no physically realizable results?
If it is indeed the latter, what impact does this have on theories which
hope to do `new physics' based on using Clifford algebra?

\section*{II. Metric Signature in Classical Electrodynamics}
Classical physics can be formulated within the language of vector
(more generally tensor) algebra/calculus.  These mathematical systems
have an explicit (or at least implicit) dependence upon the 
``metric'' associated with the geometry of the underlying physical
space.  In this section we ask: {\it Can classical electrodynamics
be formulated such that it is indpendent of the ``signature''
(absolute sign) of the metric?}  First we need to investigate
which mathematical structures are preserved under the map which
inverts the sign of the signature.  We find in particular that the
{\it Gibbs Curl} is not well behaved.  Hence for example, Faraday's law
cannot be cast in metric signature invariant form using Gibbs algebra.
However, the 3D Clifford algebra formulation allows for a signature
invariant equation, suggesting that this language is better suited
to express physical theories within which signature invariance is desired.

\subsection*{A. Algebra and Signature}
Given a set of basis vectors $\basise{i}$, the {\it metric} is defined
in terms of the inner ``dot'' product,
$$g_{ij}=\basise{i}\cdot\basise{j}=\basise{j}\cdot\basise{i}.\eqno(1a)$$
Under a change of metric signature: \tilt{g_{ij}}, it is presumed that
coordinates, differentials $dx^j$ and physical observables (e.g. charge
density $\rho$) will not change.  The dot product of two physical vectors
(e.g. electric and magnetic fields) will transform:
${\bf E}\cdot{\bf B}\rightarrow - {\bf E}\cdot{\bf B}$, while the
gradient: $\nabla=\basise{j} g^{jk}\partial_k
\rightarrow -\nabla$ must acquire a sign change.  Putting it together, 
Gauss' law: $\nabla\cdot{\bf E}=\rho$, is invariant under change of signature,
hence we call it {\it signature form invariant}.
In contrast, Faraday's law expressed in Gibbs vectors:
$\nabla\times{\bf E}=-\partial_t {\bf B}$
is \underline{not} signature invariant, an anomalous change of sign will 
appear on the left side, but not the right.
The potential equations are even more problematic when
Gibbs vectors are used,
$${\bf E}=-\nabla V +\partial_t{\bf A}, \eqno(1b)$$
$${\bf B}=\nabla\times{\bf A}. \eqno(1c)$$
In order to keep the electric field unchanged, we must have {\bf A}
invariant (and \tilt{V}), but then the magnetic field would not be invariant.

The definition of a Clifford algebra is,
$$g_{ij}={1 \over 2} \{\basise{i},\basise{j}\}=
\basise{i}\cdot\basise{j}.\eqno(2a)$$
In this paper we will always restrict ourselves to 
$g_{ij}=\pm \delta_{ij}$, an orthonormal basis.
Hence $\basise{i}\basise{j}=
-\basise{j}\basise{i}=\basise{i}\wedge\basise{j}$ for
$i \ne j$, so we can just leave out the $\wedge$ in products
of basis elements and use the compact notation
$\basise{12}=\basise{1}\basise{2}$.  In {\tt 3D} space, the products
of the the 3 basis vector generators create the 
full 8 element basis for the algebra: $\{1, \basise{1},
\basise{2},\basise{3}, \basise{12}, \basise{23},\basise{31},
\basise{123} \}$.  Note the unit trivector ${\cal I}=\basise{123}
=\basise{1}\basise{2}\basise{3}$ commutes with all the elements.
In usual Euclidean space we have $g_{ij}=\delta_{ij}$, abbreviated 
as $(+++)$ {\it signature}.  The associated Clifford algebra is
${\bf C}(2)={\tt End \ }
{\bf R}^{3,0}$, commonly known as the {\it Pauli algebra} of
2 by 2 complex matrices.  It is completely inequivalent to the
$^2{\bf H \ }={\tt End \ }{\bf R}^{0,3}$
(block diagonal quaternionic matrics) associated with the 
$(---)$ signature\cite{Porteous}.  In particular, in $(+++)$, the unit trivector
${\cal I}^2=-1$, hence behaves like the usual abstract $i$,
while in the $(---)$ signature ${\cal I}^2=+1$ and 
\underline{no} element plays the role of $i$.
Classical physics can apparently be expressed in either metric,
possibly because both have the same bivector subalgebra 
hence quadratic forms in each are invariant under the rotation
symmetry group $O(3)$.

Although the algebras are inequivalent, it is possible for algebraic
formulas to still be invariant in form.  For example, \EQN{2a} as well
as the wedge product of two vectors:
${\bf a}\wedge{\bf b}={1 \over 2} [{\bf a},
{\bf b}]$ are valid in both $(+++)$ and $(---)$.
To make general statements about signature form invariance we need
the algebraic map associated with the
replacement \tilt{g_{ij}}.
It is a common trick to let $\basise{j}\rightarrow i\basise{j}$, such that
\tilt{\basise{j}\cdot\basise{k}}, however
this is ``cheating'' as there is no $i$ in the $(---)$ algebra.
In mapping from ${\bf C}(2)\rightarrow {^2{\bf H}}$ we use the 
{\it tilt transformation} introducted by
Lounesto\cite{Lounesto93},
$$ab\rightarrow b_ea_e + b_ea_o+b_oa_e-b_oa_o, \eqno(2b)$$
where the subscripts `e' and `o' refer to the even and odd parts
respectively.  Under \EQN{2b} the Clifford definition of the dot
product of two vectors: 
${\bf a}\cdot{\bf b}={1 \over 2}\{{\bf a},{\bf b}\}$
will transform: \tilt{{\bf a}\cdot{\bf b}} as desired.
The norms of the odd elements will change sign:
$\til{{\cal I}^2}, \til{\basise{j}^2}$, while the even elements 
(e.g. bivectors) will not.  While the basis elements such as
${\cal I}=\basise{123}$ map unchanged in form, the form of the
relation between a bivector and the dual of a vector (in 3D) changes:
\tilt{{\cal I}{\bf a}}.
One can now construct a nonstandard definition of the Gibbs cross product
which is metric signature form invariant:
$${\bf C}={\bf A}\times{\bf B}=-{\cal I}{\bf A}\wedge
{\bf B}. \eqno(2c)$$
The Clifford algebra analogies of
Faraday's law and \EQN{1b},
$$\nabla\wedge{\bf E}=-{\cal I}\partial_t{\bf B}, \eqno(2d)$$
$${\cal I}{\bf B}=\nabla\wedge{\bf A}, \eqno(2e)$$
are completely signature invariant whereas the Gibb's vector forms
were not!  Apparently Clifford algebra 
is a better language in which to express principles of physics
which should be metric signature invariant.

\subsection*{B.  Nonrelativistic Electrodynamics}
What conditions would insure a metric signature invariant formulation
of non-relativistic electrodynamics?  Let us review how the
metaprinciple of isotropy (no preferred direction) is imposed.
Equations of motion are derived from a Lagrangian (via the principle
of least action).  It is sufficient to require the Lagrangian
to be invariant under the rotation group $O(3)$.  The equations
of motion derived from such a Lagrangian will be rotationally
form invariant.

As an example, let us consider the Lagrangian for 
a non-relativistic charged particle
in a classical electromagnetic field,
$${\cal L}={1 \over 2} mv^kv^jg_{jk} + e v^jA^kg_{jk}-eV+
{1 \over 2}(E^jE^k-B^jB^k)g_{jk} , \eqno(3a)$$
$$E^k=-g^{kj}\partial_jV+\partial_tA^k, \eqno(3b)$$
$$B^k=g\ g_{jn}\epsilon^{mjk}\partial_mA^n, \eqno(3c)$$
where the factor of $g=det(g_{jn})$ in \EQN{3c} is needed to overcome the
problems with the Gibbs curl mentioned in the previous section in
\EQN{1b}.
Since the terms are all quadratic in the vectors, the Lagrangian is
invariant under the rotation group $O(3)$.

Now lets consider if this Lagrangian is invariant under the {\it signature
transformation}.  Under the replacement of
$g_{jk}\rightarrow -g_{jk}$ we see that we must also reflect the
scalar potential
$V\rightarrow -V$ in order for the electric field \EQN{3b} to be
invariant.  The Lagrangian is not really invariant, but transforms
\tilt{\cal L}, acquiring an overall minus sign which does not
change the equations of motion.  Hence the Lagrangian \EQN{3a}
will generate a signature invariant form of electrodyamics as
desired.

\subsection*{C.  Relativistic Electrodynamics}
In relativity, the metaprinciple of isotropy  is generalized
to include time as the fourth dimension.  Either \R4 or \H2 
signatures can be used; apparently neither special nor general
relativity can distinguish between the two signatures.  Perhaps
this is because relativity is only concerned with the group
structure $SL(2,C)=SO(1,3)=SO(3,1)$, not from where the group
derives\cite{Voros}.

In the \R4 signature, the {\it proper time} is defined,
$$d \Tau^2 =-g_{\mu\nu}dx^\mu dx^\nu, \eqno(4)$$
which is a {\it scalar}, the same value in
all reference frames connected by a Lorentz transformation.  The
{\it four-velocity} is defined in terms of this invariant,
$$u^\alpha={{dx^\alpha} \over {d\Tau}}, \eqno(5a)$$
$$u^\alpha u^\beta g_{\alpha\beta}=-1. \eqno(5b)$$
Under the transforamtion \tilt{g_{\mu\nu}}, \EQN{4} is not invariant,
unless we propose: $d\Tau\rightarrow id\Tau$.  We are perhaps
cheating abit here by introducing an abstract $i$, but it will
completely vanish in the end.
It follows that the four-velocity definition transforms:
$u^\alpha \rightarrow -i u^\alpha$, such that \EQN{5b} is
``signature form invariant''.

In the Lorentz gauge, Maxwell's equations in the \R4 signature become,
$$g^{\alpha\beta}\partial_\alpha\partial_\beta A^\nu=J^\nu. \eqno(6a)$$
If we wish the current to be invariant under the signature change
\tilt{g_{\mu\nu}}, then the potential must transform: \tilt{A^\mu},
which is different than we had for the non-relativistic case.
Consider now the relativistic electrodynamic Lagrangian
in the \R4 signature,
$${\cal L}=m\sqrt{-g_{\alpha\beta}u^\alpha u^\beta }
+ e A^\alpha u^\beta g_{\alpha\beta}. \eqno(6b)$$
Under the signature change, the Lagrangian will transform:
${\cal L}\rightarrow -i{\cal L}$.  The factor of $i$ is a surprise,
however since the proper time transforms 
$d\tau \rightarrow id\tau$, the
{\it action} ${\cal A}=\int{{\cal L} d\tau}$ will be invariant.
Again, signature invariant equations of motion will be generated.

\section*{III.  Spin Space and Spin Metrics}
Quantum physics requires the introduction of a new space associated
with the spin degrees of freedom of a wavefunction.  In order for
physical observables to be invariant under the change of signature
of real space, it is found that spin space must transform non-trivially.
Further it is 
found that Dirac theory cannot be written in signature invariant
form within standard formalism.

\subsection*{A.  Dirac Algebra}
Historically, the Dirac equation was derived by factoring the
Klein-Gordon operator.  In the \R4 signature,
$$(\Box^2-m^2)=(\BOXY-m)(\BOXY+m), \eqno(7a)$$
where $\BOXY=\gamma^\mu\partial_\mu$ and
$\BOXY\BOXY=\Box^2=g^{\mu\nu}\partial_\mu\partial_\nu$
is the D'Alembertian.  The Dirac matrices $\gamma^\mu$ are
related to the metric,
$$2g^{\mu\nu}=\{\gamma^\mu ,\gamma^\nu \}=
\gamma^{\mu A}_{\ \ \ B}\ \gamma^{\mu B}_{\ \ \ A}, \eqno(7b)$$
where the indices ``A'' and ``B'' of $\gamma^{\mu A}_{\ \ \ B}$ refer to
the row and column of the matrix $\gamma^\mu$.

Standard Dirac algebra is ${\bf C}(4)$, meaning
$4\times4$ {\it complex} matrices\cite{Porteous}.  This
corresponds to a {\tt 5D} Clifford algebra where
$i=\basis{12345}=\basis{1}\basis{2}\basis{3}\basis{4}\basis{5}$
is the unit 5-volume.  This algebra admits three possible {\tt 5D}
metric signatures: $(-----), (+++-+)$ and $(---++)$.
The latter two of these show that both {\tt 4D} signatures of
\R4 and \H2 are contained as subalgebras of {\bf C}(4).
Choosing one {\tt 4D} metric signature over the other is
simply taking a different ``slice'' of global {\tt 5D} space.

The {\tt 5D} signature change is really
perhaps only a reshuffling of $(+++-+)\rightarrow(---++)$ which
is all still in the same ${\bf C}(4)$ algebra.
According to \EQN{7b}, under a signature change \tilt{g_{\mu\nu}}
we must have something like \tilti{\gamma^\mu} (except
$\gamma^5$ is invariant).
This is actually a {\tt 5D} duality transformation, trading
vectors for their dual quadvectors.
Perhaps the metaprinciple at work here is something like:
{\it The laws of physics are invariant under a global 5D
duality transformation}.
  
\subsection*{B.  Wavefunctions}
It is easy to show that wavefunctions cannot be invariant under
the signature change.  In order for the momentum density,
$$p^\mu = -{i \hbar \over 2}[\Psi^\dagger (\partial_\nu \Psi)
-(\partial_\nu \Psi^\dagger)\Psi]g^{\mu\nu}, \eqno(8a)$$
to be invariant under \tilt{g^{\mu\nu}}, we must have something
like $\Psi \rightarrow \Psi^*$.  The situation is more complicated
when spin degrees of freedom are included.  The solution $\psi^A$
to the Dirac equation is a 4 complex component column bispinor.
In the \R4 signature the observable Dirac current is bar invariant,
$$j^\mu=i\BAR{\Psi}\  \gamma^\mu\ \Psi=
\Psi^\dagger\gamma^4\gamma^\mu\Psi, \eqno(8b)$$
where $\BAR{\gamma}^\mu=-\gamma^\mu$
and $\BAR{\Psi}=\Psi^\dagger i \gamma^4$ in the standard matrix 
representation. 
Under the signature change \tilt{g_{\mu\nu}}, the current 
$j^\mu$ should be invariant,
and remain bar invariant.  In the \H2 signature the current should
have the form:
$j^\mu=\BAR{\Psi}\ \gamma^\mu\ \Psi$ where, contrary to the 
sensibilities of algebraists, the definition of the
bar is different: 
$\BAR{\gamma}^\mu=+\gamma^\mu$ and $\BAR{\Psi}=\Psi^\dagger \gamma^4$.
The problem is to choose the map for $\Psi$ such that \EQN{8a} will
be signature invariant, as well as the Dirac current \EQN{8b}. 
One possibility is,
$$\Psi \rightarrow i\gamma^2 \Psi^*, \eqno(9)$$
which corresponds to the charge conjugation operator.  Note that under
\EQN{9} the norm transforms,
$$\BAR{\Psi}\Psi=|\psi^1|^2+|\psi^2|^2-|\psi^3|^2-|\psi^4|^2
\rightarrow -(\BAR{\Psi}\Psi). \eqno(10)$$
A possible physical interpretation would be that there is a
connection between signature change and charge conjugation
symmetries.

\subsection*{C.  Wave Equation}
It is immediately clear that one cannot write a signature invariant
Klein-Gordon equation.  Consider that \EQN{7a} in the opposite 
signature of \H2 has the form,
$$(\Box^2+m^2)=(\BOXY+i\ m)(\BOXY-i\ m), \eqno(11)$$
which differs from eq. (7a) by the factors of $i$.
Since the Dirac equation is based on this factorization, it follows 
that one cannot write a signature invariant form of the Dirac
equation (nor a Lagrangian) in standard notation.

\section*{IV. Multivector Quantum Mechanics and Signature}
Multivector Quantum Mechanics formulates wave equations entirely 
within the real Clifford algebra of {\tt 4D} spacetime.  There is no
spin space separate from real space; a spinor wavefunction is now
represented by a {\it multivector} (an aggregate of scalar, vector,
bivector, etc.).  Under a signature change, the wave function
must hence transform by the same rule \EQN{2b} as the underlying geometry
of spacetime; it cannot have its own rule like \EQN{9}.
Unlike standard theory, one can now formulate a signature invariant Dirac-like
equation under certain restrictions.  Further, under the tilt
transformation the generalized multivector equation suggests a 
new interchange symmetry between spin and isopsin.

\subsection*{A. The Algebraic Dirac Equation}
The Clifford algebra associated with \R4
signature is: ${\bf R}(4)$,
isomorphic to real $4\times4$ matrices\cite{Porteous},
otherwise known as the 
Majorana algebra.  It is inequivalent to the ${\bf H}(2)$ algebra
($2\times2$ matrices with quaternionic entries) associated with
metric \H2.  While both algebras have 16 basis elements, they have
very different properties and substructures.  In the 
{\bf R}(4) algebra there are ten elements:
${\bf R}_+=\{1, \bas{1}, \bas{2},\bas{3},
\bas{41},\bas{42},\bas{43},
\bas{412},\bas{423},\bas{431}\}$ for which if $\Gamma \in
{\bf R}_+$ then $\Gamma^2=+1$, while for
$\Gamma \in {\bf R}_{-}=\{\bas{4},\bas{123},\bas{1234},
\bas{12},\bas{23},\bas{31} \}$ we have $\Gamma^2=-1$.
In the {\bf H}(2) algebra however we have a set of 6 elements with
positive squares: ${\bf H}_+=\{\bas{1},\bas{4},\bas{14},
\bas{24},\bas{34},\bas{123}\}$ and negative squares for the remaining
10 elements: ${\bf H}_{-}=\{\bas{1},\bas{2},\bas{3},
\bas{12},\bas{23},\bas{31},\bas{412},\bas{423},\bas{431},
\bas{1234} \}$.
Nethertheless, the even subalgebra of both of these
algebras contain the same (special) Lorentz group $SL(2,C)$.

Clifford algebras associated with even dimensional spaces 
(such as {\tt 4D}) always
lack a global commuting $i$, regardless of metric signature.
Hence one cannot do the factorization of \EQN{11} within the
Clifford algebra associated with \H2 space.
This has led many authors to argue that the \R4 signature is the
only ``correct'' model of spacetime because the factorization of 
\EQN{7a} is possible.  In this viewpoint, there is an absolute
signature of the universe.
However, one of the new freedoms we have in these
theories is the ability (indeed at times necessity) of
{\it dextrad multiplication}\cite{Pezz692},
i.e. operation on the right side of the wavefunction.
Regardless of the metric signature, consider the generalized form,
$$\Box\Psi=m\Psi\Gamma, \eqno(12a)$$
$$\Psi=\Box\phi+m\phi\Gamma, \eqno(12b)$$
where $\Gamma$ is some basis element of the Clifford algebra.
Substituting \EQN{12b} into \EQN{12a}, we recover the
Klein-Gordon equation,
$$\Box^2\phi=m^2\phi\Gamma^2. \eqno(12c)$$
For the \R4 signature choose
$\Gamma \in {\bf R}_+$, while in the \H2 case choose $\Gamma \in
{\bf H}_{-}$.  In both situations there are 10 choices.  
One can only construct a {\it signature invariant
formulation} by choosing one of the 6 elements which is the intersection
of these two dissimilar sets: $\Gamma \in {\bf R}_+ \cap {\bf H}_- 
=\{\bas{1},\bas{2},
\bas{3},\bas{412},\bas{423},\bas{431}\}$, all which are odd
geometry.

The inclusion of electromagnetic interactions requires some new element
$\Lambda^2=-1$ to play the role of the $U(1)$ gauge generator.  Since
no element in {\tt 4D} will commute with all four $\gamma_\mu$, this
element must be applied ``dextrally'' (on the right side).  The
generalization of \EQN{12a} is,
$$\Box \Psi - m\Psi \Gamma - e A^\mu \gamma_\mu
\Psi \Lambda=0,\eqno(13a)$$
$$[\Lambda,\Gamma]=0. \eqno(13b)$$
Although constrained by \EQN{13b}, the choice of $\Lambda$ is not
unique.  It is up to the author to show that given his preselection
of metric signature, his choice of $\Gamma$ and $\Lambda$ will allow
for a wavefunction solution with 8 degrees of freedom which
is isomorphic to standard Dirac theory (e.g. has proper parity,
charge conjugation and Lorentz transformation properties).  In the
\H2 signature, 
Hestenes\cite{Hestenes} favors $\Lambda=\bas{12}$ with
$\Gamma=\bas{412}$.  Lounesto\cite{Lounesto93} points out
that the same choices will work in the opposite {\it tilted}
\R4 signature.  
However, as we shall see in the next section, it does \underline{not}
necessarily follow that the formulation is {\it signature
form covariant} under the tilt transformation \EQN{2b}.

\subsection*{B.  Weak Tilt Covariance and the Dirac Equation}
Unfortunately, there is no general agreement on the proper 
generalization of a Lagrangian for multivector quantum mechanics.
We choose a form which reduces to the standard in the \R4
signature for $\Gamma=1$ and is
also generally Dirac-bar invariant,
$${\cal L}=\BAR{\Psi}(\Box\Psi)-(\BAR{\Psi}\Box)\Psi
-m\{\BAR{\Psi}\Psi,\Gamma\}
+\{\BAR{\Psi}\basis{\mu}\Psi,\Lambda\}A_{\mu}.  \eqno(14)$$
The constraint $\BAR{\cal L}={\cal L}$ insures invariance under
the Lorentz group, at the cost of requiring
$\BAR{\Gamma}=+\Gamma$ and $\BAR{\Lambda}=-\Lambda$.  In either
metric signature this restricts $\Gamma$ to 4 choices; if one
further insists on signature invariance it limits one to
$\Gamma \in \{\bas{412},\bas{423},\bas{431}\}$.  Having made
a selection for $\Gamma$, the choice of $\Lambda$ is fixed, e.g.
if $\Gamma=\bas{412}$ one must have $\Lambda=\bas{12}$ in
order to satisfy \EQN{13b}.  Different from standard theory,
in real Clifford algebra we have,
$\BAR{\gamma}_\mu =-\gamma_\mu$ in either metric.

Regardless of these restrictions, we find that \EQN{13a} can have 
the same form in both metric signatures.  However it does not necessarily
follow that \EQN{13a} will be form invariant under the tilt 
transformation.  In fact, the application of \EQN{2b} to (the bar
of) \EQN{13a} returns the same \EQN{13a}, except that the
wavefunction transforms: $\Psi \rightarrow \widetilde{\Psi}
=-\bas{1234}\BAR{\Psi}\bas{1234}$.  Full signature invariance
can be obtained only if the wavefunction is restricted such that
$\Psi = \pm \widetilde{\Psi}$.  The choice of $\Psi=+\widetilde{\Psi}$
limits the solution to only 6 components, is too restrictive
for Dirac theory (which requires 8 degrees of freedom).  The other
choice $\Psi=-\widetilde{\Psi}$ has 10 components, but the 6 odd
geometric parts are completely decoupled from the 4 even parts if
$\Gamma$ is odd geometry as previously argued.  Dirac theory 
necessarily has all 8 components coupled.  {\it Therefore we 
\underline{cannot} have
a fully signature \underline{invariant} Dirac equation under the tilt
transformation of \EQN{2b}}.

This is consistant with the arguments presented in section III.B, 
showing that wavefunctions (in standard theory) could not be
invariant under a signature change.  A weaker condition would be
to see if the physical observables are invariant under the tilt
transformation, allowing $\Psi \rightarrow \widetilde{\Psi}$.
We shall call this {\it Weak Tilt Covariance}.
Let us consider a Greider-Ross\cite{Greider} {\it multivector current}
which for \EQN{13a} obeys a generalized conservation equation,
$$\partial_\mu j^\mu =m\  [\ \BAR{\Psi}\Psi,\Gamma\ ], \eqno(15a)$$
$$j^\mu=\BAR{\Psi}\gamma^\mu\Psi=-\gamma_4\Psi^\dagger
\gamma_4\gamma^\mu\Psi. \eqno(15b)$$
In contrast to \EQN{10}, the even and odd multivector parts 
of quadratic forms unfortunately do not transform the same
under the tilt transformation \EQN{2b},
$$(\BAR{\Psi}\Psi)_o \rightarrow(\Psi\BAR{\Psi})_o. \eqno(16a)$$
$$(\BAR{\Psi}\Psi)_e \rightarrow(\Psi\widetilde{\Psi})_e
\ne (\Psi\BAR{\Psi})_e. \eqno(16b)$$
The conservation law \EQN{15a} can be made invariant under the tilt
only if the wavefunction is restricted to be either pure even or pure
odd geometry and transforms $\Psi \rightarrow 
\pm \widetilde{\Psi}$.  A pure even (or odd) wavefunction will have
8 degrees of freedom, therefore is sufficient to describe
Dirac theory.  Under these restrictions, noting that
$\BAR{\Psi}_e=+\widetilde{\Psi}_e$ and $\BAR{\Psi}_o=
-\widetilde{\Psi}_o$, the Lagrangian \EQN{14}
is found to be invariant under the tilt transformation, insuring 
that the ``physics'' will be independent of signature choice.

\subsection*{C. Generalized Tilt Covariant Gauge Interactions}
The lack of unique choice for $\Gamma, \Lambda$ in \EQN{13a}
has two sources, generally
unrecognized by authors.  First, the full multivector solution
$\Psi$ to \EQN{13a} has 16 degrees of freedom, while only 8 are
needed to describe standard Dirac theory (i.e. isomorphic to
a 4 complex component bispinor).  Some author's choices for 
$\Gamma, \Lambda$ will only represent Dirac theory 
if the wavefunction is restricted by some criteria
to a specific 8 geometric components (n.b.\ Hestenes\cite{Hestenes} and
Lounesto\cite{Lounesto93} require it to be pure odd or pure even).
In contrast, it has been argued\cite{Pezz592}
that the full multivector solution of \EQN{13a} represents an
isospin doublet of bispinors, for which \underline{both}
of the isospin components have the correct electromagnetic
interaction only for $\Gamma=1, \Lambda=\gamma_4$ in \R4.
In the previous section, the choice of $\Gamma=1$ was excluded
because it would work only in the \R4 signature.
However, there is a counterpart in the \H2 signature which can
be found by applying
\EQN{2b} to \EQN{13a},
$$\Box \Phi-m\widehat{\Phi}
-A^\mu\gamma_\mu\Phi\gamma_4=0, \eqno(17a)$$
$$\widehat{\Phi}=-\gamma_{1234}\Phi\gamma_{1234}, \eqno(17b)$$
where $\Phi=\widetilde{\Psi}$.  The mass term now involves the
{\it grade involution} of \EQN{17b}, a form not considered in
\EQN{12a}.   Indeed,
Lounesto\cite{Lounesto93} obtained a similar result and was
concerned about the lack of physical interpretation for the
appearence of a grade involution.

This leads us to our second main point about the multitude of
choices for $\Gamma$ and $\Lambda$ in \EQN{13a}.
Consider that the electromagnetic interaction term
$(\gamma_\mu \Psi \Lambda)$ involved both left and right sided
multiplication.  Multivector theory allows for a variety of new
gauge interactions base on this {\it bilateral} (two-sided)
form[6,7].  We have proposed the generalized
equation\cite{PezzMex}, which includes all possible couplings,
$$\Box\Psi=-\Ed{i}\Psi {\bf F}_{(j)}\Omega^{(ij)}.\eqno(18)$$
The factor $\Ed{i}$ (or ${\bf F}_{(j)}$) is one of the
16 basis elements of the
Clifford algebra.  For example let: $\Ed{0}=1, \Ed{\mu}=\gamma_\mu$
\ (for $\mu=1,2,3,4$),
and $\Ed{15}=\epsilon=\gamma_{1234}$.  The grade involution of
\EQN{17b} is simply a special case where the mass is identified
with $\Omega^{(15,15)}$.  The {\it bilateral
connection} $\Omega^{(ij)}$ is a generalized set of gauge currents,
subject to a Lagrangian constraint that non-zero components must have
corresponding factors of $\Ed{i}, {\bf F}_{(j)}$ both either bar 
negative or bar positive.  Explicitly we rewrite \EQN{18},
$$\Box \Psi = \Psi (m_1 + \g5\bas{\mu}a_0^{\ \mu}+\g5{\cal X})
+\g5\Psi(\eta + \g5\bas{\mu}\pi^\mu+\g5 m_2)$$
$$+\basis{\mu}\Psi(\bas{\nu} b_\mu^{\ \nu}+\bas{4}\bas{A} a_\mu^{\ A}
+\g5\bas{4}\bas{A}\rho_\mu^{\ A})
+\g5\basis{\mu}\Psi(f_{1\mu}+\g5\bas{\nu}a_{1\mu}^{\ \ \nu} +\g5 \phi_\mu)$$
$$+\basis{\alpha\beta}\Psi(\bas{\mu}S_{\alpha\beta}^{\ \ \ \mu}
+\bas{\mu\nu}R_{\alpha\beta}^{\ \ \ \mu\nu}), \eqno(19)$$
where $A=1,2,3$ and $\g5=\gamma_{1234}$.
The symbology of the gauge fields in \EQN{19} is deliberate,
reflecting that nearly all the light unflavored mesonic interactions
can be accomodated in the above scheme\cite{Meson}.  For full
$SU(2)\times U(1)$ electroweak theory in \R4, the correct 
correspondence has been demonstrated as\cite{Pezz592}:
mass $m=m_1$, electromagnetism 
$A_\mu=b_\mu^{\ 4}$ (generator $\gamma_4$), and vector bosons 
$W^A_\mu = \rho_\mu^{\ A}$; in particular $Z=W^0$ has generator
$\gamma_{12}$ which commutes with $\gamma_4$.  On the other hand,
the tilted form of the Hestenes-Dirac equation presented by
Lounesto\cite{Lounesto93} for \R4 has:
$m=a_0^{\ 3}$, and $A_\mu=\rho_\mu^{\ 3}$.  Comparing with above, it looks like
they are using the Z boson for electromagnetism.  Indeed as long as
you restrict the wavefunction to a single (8 degree of freedom)
Dirac bispinor, which is an isospin eigenstate, you cannot 
distinguish between the two.  However this choice of 
$\gamma_{12}$ for the generator of electromagnetism cannot
accomodate a full electroweak theory.  There is no set of 
elements which will commute with this and $\Gamma=\gamma_{412}$
that have the necessary $SU(2)$ group structure.
 
Our generalized \EQN{18} is {\it manifestly weak tilt covariant}.
Application of \EQN{2b} to \EQN{18} is reduced to a 
transformation among the various
components $\Omega^{(i,j)}$.  Explicitly, in terms of \EQN{19} we find
under change of signature with $\Psi\rightarrow\widetilde{\Psi}$
the gauge fields transform,
$$m_1 \leftrightarrow -m_2, \eqno(20a)$$
$$b_\mu^{\ \nu}\leftrightarrow -a_{1\mu}^{\ \ \nu}, \eqno(20b)$$
$$R^{\alpha\beta\kappa\lambda} 
\leftrightarrow -\epsilon^{\alpha\beta\rho\sigma}
\epsilon^{\kappa\lambda\iota\theta}R_{\rho\sigma\iota\theta}, \eqno(20c)$$
while the rest are invariant.  It is thought that some of these symmetries
can be interpreted in terms of why we only see an electric monopole OR a
magnetic monopole both, and a new symmetry between isospin and regular spin.
A more complete treatment will show how the conserved currents map from
one system to the other (hence the form of the observables different).

\section*{V. Summary and Conclusions}
Classical mechanics (electrodynamics and geometrodynamics) does indeed
appear to be unable to determine absolute metric signature.  The formulations
can be easily cast in signature invariant form.  The situation is quantum
mechanics is less clear.  While the wave equation may not be signature
form invariant, the observables can be put into such a form.  The spin 
metric's signature does not necessarily change with the space metric
signature, nor would any signature associated with isospin space.

However, in multivector wavemechanics, spin space, indeed even isospin
space is described in one single {\tt 4D} spacetime algebra.  Therefore
the ``spin basis'' structure cannot be isolated from changes in the
spacetime signature because of the inequivalence of the \R4 and \H2
Clifford algebras.  While at first it appears that the multivector
formulations are not at all similar in the different signatures, this is
because most authors only consider a subset of the full problem.  Hence
each is slicing a different way through a much richer reality, only seeing
some features, e.g. only the electromagnetic interactions but not the
weak interactions.  We have proposed the most general form of
the multivector wave equation, which includes all other author's works as
a subset.  This equation does show full signature form invariance, when
one considers all possible gauge interactions and all 16 components of
the multivector wavefunction.

\end{document}